\documentclass[11pt,twoside]{article}
\usepackage{asp2010}

\resetcounters

\begin{document}

\title{Direct modeling of neutral helium in the heliosphere}
\author{Hans-Reinhard~M\"uller$^{1,2}$
\affil{$^1$Department of Physics and Astronomy, Dartmouth College,
Hanover NH 03755, USA}
\affil{$^2$Center for Space Plasma and
Aeronomic Research, University of Alabama, Huntsville AL 35805,
USA} }

\begin{abstract}
Several years of neutral particle measurements by the NASA/IBEX
mission have yielded direct observations of interstellar neutral
helium and oxygen. The data indicate the presence of secondary
neutral helium and oxygen, which are created within the
heliosphere by charge exchange involving helium or oxygen ions.
This contribution describes a detailed conserving calculation
method based on Keplerian orbits that has been developed to
characterize helium distribution functions throughout the
heliosphere, in particular in the innermost heliosphere, while
accounting for loss and production of neutral particles along
their path. Coupled with global heliosphere models of plasma
distributions, this code is useful for predicting the fluxes of
heavy neutral atoms at spacecraft detectors, so enabling
inferences on the characteristics of the interstellar medium.
\end{abstract}

\section{Introduction}
On its galactic journey, the Sun traverses a cloud of interstellar
gas that is moderately warm and dense, and is partially ionized,
meaning that it also contains neutral atoms (dominantly hydrogen,
but also helium, oxygen, and others). While the interstellar
plasma interacts with the solar wind to form the global
heliosphere, interstellar neutrals are unaffected and proceed into
the heliosphere in their original velocity state. As such they are
direct messengers of the interstellar medium, detectable by
spacecraft in the inner solar system \citep[like NASA/IBEX;
][]{Bzowski12, Moebius12}. This has intensified interest in
accurate heliospheric neutral modeling, including secondaries.

The entrance of heavy interstellar neutrals into the heliosphere
has been modeled in detail in the past \citep[e.g.,][and
references therein]{Mueller04, Izmodenov04, Bzowski12}. These
calculations are typically variants of models and calculations of
neutral interstellar hydrogen in the heliosphere, which have a
long history spanning four decades \cite[for reviews,
see][]{Axford72, Zank99, Izmodenov06}. Simulations treat the
heavies as test particles on the background of the global
hydrogen/proton heliosphere. Monte-Carlo methods are useful in the
global heliosphere, but statistics are poorer in the innermost
heliosphere where IBEX measures. Reverse-trajectory methods are
therefore popular to calculate heavy neutrals in the heliosphere,
but typically the input flux is approximated at an outer boundary
that is still well inside the heliosphere.      This paper focuses
on a method combining the best of both approaches, starting from
the pristine interstellar medium, but calculating the phase space
distribution function (PDF) at an arbitrary point of interest in a
conserved way through reverse trajectories while allowing for
arbitrary ionization (loss) and production terms.

\section{Scientific framework and numerical details}

In a heliocentric frame of reference, the motion of the neutrals
is governed by the central gravitational potential of the
Sun, such that the neutrals are on Keplerian trajectories, their
paths deflected close to the Sun. The potential is weak for
hydrogen which experiences an additional outward radiation
pressure force that balances most of gravity. During solar maximum
when radiation levels are highest in the solar cycle, radiation
even overcompensates gravity, leading to an effectively repulsive
potential for hydrogen during these times. Heavy neutrals are
easier to handle, and the unimportance of radiation pressure for
them means that the corresponding central potential is
conservative and time-independent.

Primary (interstellar) particles are on open trajectories
(hyperbolae, and possibly parabolae as limiting case). Each
individual trajectory can be described geometrically by specifying
the orbital plane, the direction of perihelion, and the orbital
eccentricity $e$. If $\theta$, the angle that a particle's
position vector makes with the direction of perihelion, is given,
then the particle's radial distance and velocity are determined
with simple formulae from celestial mechanics ($r(\theta)$;
$v_r(\theta)$; $v_{\theta}(\theta)$); the time to or since
perihelion $t(\theta)$ can also be obtained relatively easy albeit
not in closed form. If any of the latter quantities is given
instead, the other variables are determined just as easy from it,
even if sometimes two solutions arise describing situations which
are symmetric with respect to the perihelion (e.g., $\theta(r)$).
Each individual trajectory can also alternatively be described
physically by specifying conserved quantities including total
specific energy $E$, specific angular momentum $\vec l$, and the
eccentricity vector $\vec a$ \citep{Mueller12}. Again with
trajectories defined in this manner, any one given value for
angle, distance, velocity or time will yield all the respective
others. There are of course numerous relations that link orbital
elements to the physical quantities (for instance, the orbital
plane is perpendicular to $\vec l$, and $e$ and the perihelion
distance $r_{min}$ depend only on $l$ and $E$).

On their way through the heliosphere, atoms have a probability of
being lost. In the case of interstellar helium, in order of
importance for helium atoms detected at a 1 AU distance (e.g.,
location of IBEX), the processes are loss due to photoionization,
loss due to charge exchange (c.x.) with slow helium ions in the
outer heliosheath, double-c.x.\ with solar wind alpha particles,
and c.x.\ with solar wind protons \citep{Bzowski12}.      Earlier
studies \citep[e.g.,][]{Mueller04} were limited in admitting only
c.x.\ with protons as a loss process. Secondary helium is created
as the neutral product of a c.x.\ collision between a helium ion
and a neutral, the dominant interaction being bow-shock
decelerated interstellar He$^+$ exchanging charge with
interstellar helium in the outer heliosheath (loss of a primary
particle and at the same time creation of a secondary neutral in a
different velocity state, reflecting the underlying plasma
distribution function). Of course, once created, a secondary
neutral on its further trajectory suffers the same types of losses
as the primary neutrals.

Two frames of reference are most appropriate for the neutral
calculations: (1) The ``ISM frame" in which the $x$ axis points
antiparallel to the interstellar (ISM) flow so that the ISM flow
vector at infinity is $(u_{x,ISM}, 0, 0)$ with $u_{x,ISM}= -26.3$
km/s \citep{Witte04}; and (2) the ``Kepler frame" in which the $x$
axis points to the perihelion of the trajectory in question, and
the $x$-$y$ plane is the orbital plane. For each trajectory,
$a_z=l_x=l_y=0$ in its Kepler frame, leading to a convenient
reduction of conserved quantities that need to be considered.
Rotation transformations relate ISM frame coordinates to Kepler
frame coordinates, and also to other coordinates of choice such as
a heliocentric frame of reference associated with ecliptic
longitude and latitude. Spherical polar coordinates like the
latter are typically used for the heliosphere plasma background
model. $r$ is invariant under these transformations.

\begin{figure}
\centering
\includegraphics[width=0.89\textwidth]{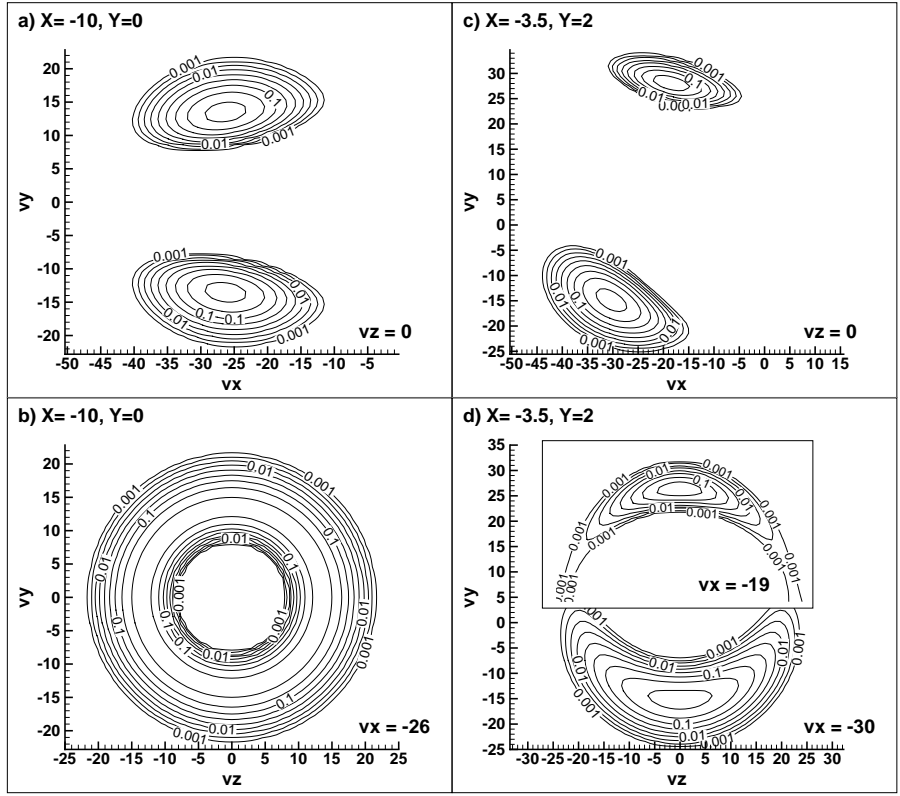}
\caption{Phase space densities of primary neutral interstellar
helium at downwind locations, on the symmetry axis ((-10.0, 0.0),
panels (a) and (b)), and at a location slightly off the symmetry
axis at (-3.5, 2.0), distance 4.0 AU (panels (c) and (d)). Both
locations are in the helium focusing cone, with the PSD being a
deformed torus loosely perpendicular to the $x$-axis; see text for
details. Presented for each PSD are two perpendicular cuts through
the 3D object, with the cut plane specified (panel (d) and its
inset have different cut planes, as indicated). The maximum PSD
value in the ISM is normalized to 1 (logarithmic contour levels
with 3 lines per dex. }
\end{figure}

The strategy to calculate a phase space distribution function
(PDF) at a point of interest $\vec r_0$ is essentially to
investigate all possible trajectories that pass through $\vec
r_0$. For the primary PDF some of these trajectories have to be
disregarded because they are not connected to the ISM (everything
outside of the heliospheric bow shock); similarly, some
trajectories that do not intersect with the outer heliosheath do
not contribute towards the secondary PDF. In principle, all these
trajectories passing through $\vec r_0$ can be obtained by
scanning the entire 3D velocity space; for each individual phase
space location $(\vec r_0, \vec v_0)$, the reverse trajectory is
uniquely determined and the velocity at infinity, or at any point
where the trajectory intersects the outer heliosheath, is
calculated in one algebraic step, while preserving the conserved
quantities by design.

Neglecting secondary neutrals for the moment, the prescription for
the primary PDF is then as follows: By Liouville's theorem, the
raw phase space density at $(\vec r_0, \vec v_0)$ is the same as
the value of the ISM Maxwellian at the calculated velocity at
infinity. Attenuation factors arise due to photoionization and
charge exchange; the photoionization factor can be cast into an
analytic form involving essentially only expressions for $\theta$
and for the asymptotic $\theta_{\infty}$ at infinity, the latter a
function of $e$. As we want to take the plasma background from an
external global heliosphere model, the c.x.\ attenuation requires
a c.x.\ loss integration along the trajectory from infinity (in
reality, from the bow shock crossing) to $\vec r_0$, which is
carried out at the $r$-grid nodes of the background model, with
$\theta(r)$ determined algebraically (conserving by design) and
the transformation matrix between Kepler and global heliosphere
frames being constant during this calculation.

In practice, instead of 3D scanning, the locations of local maxima
in the PDF are predicted algebraically: There are typically two
maxima associated with the primary neutrals, called direct- and
indirect-path velocities, where the indirect path is longer and
leads the particle closer to the Sun \citep[e.g., ][]{Mueller12}.
These velocities can be obtained in one step in the ISM frame by
requiring the velocity at infinity to be $u_{x,ISM}$, which is the
location of the maximum of the ISM Maxwellian:
\begin{equation}
(v_x, v_y) = (u_{x,ISM},0) - {f_{\mu}\over r_0 l_z}(y, r_0-x) \, ;
 \,\,\, l_z =
-\frac{y\, u_{x,ISM}}{2} \left(1 \pm  \sqrt{1+4 {f_{\mu}\over
u_{x,ISM}^2} {r_0-x\over y^2}}\right)
\end{equation}
with $(x,y)\equiv\vec r_0$, $f_{\mu}$ the constant of gravity
multiplied by the solar mass, and the upper sign the direct path.
The derivation typically makes use of the conserved quantities;
similar results have been summarized by \citet{Axford72} and
others.

The simplest practical strategy for the primary helium PDF is
hence to obtain the location of the two maxima with equation (1),
and then scan velocity space around them until the phase space
density has dropped off by, say, 4 orders of magnitude which is
equivalent to $\sim$4 thermal velocities, beyond which it can
safely be neglected. A more sophisticated procedure is to develop
the equivalent of eq.\ (1) for the fixed ISM frame with arbitrary
interstellar vectors $\vec u$ and with it, map the locus of the
edge of the ISM Maxwellian to velocities at $\vec r_0$. This gives
a closed velocity surface at $\vec r_0$ within which the
non-negligible PDF resides.

\section{Primary helium results}

Figure 1 shows two examples of applying the code, using typical
ISM characteristics \citep{Witte04}, a typical heliospheric
background model \citep{Mueller06}, and typical photoionization
and c.x.\ rates.         In the helium focusing cone on the
downwind symmetry axis (Fig.\ 1a, b), the primary helium PSD is
degenerate, with all planes containing the symmetry axis being
equally preferred orbital planes for the interstellar particles.
This leads to a torus-like PDF, with the torus perpendicular to
the $v_x$ axis and the cross section of the torus slightly
gravitationally deformed away from a circular shape. The locus of
maxima (64\% of the interstellar Maxwellian maximum due to losses)
is a circle with radius $\sqrt{-2f_{\mu}/x}=13.3$ km/s at a
constant $v_x = u_{x,ISM}$.

\begin{figure}
\centering
\includegraphics[width=0.3\textwidth]{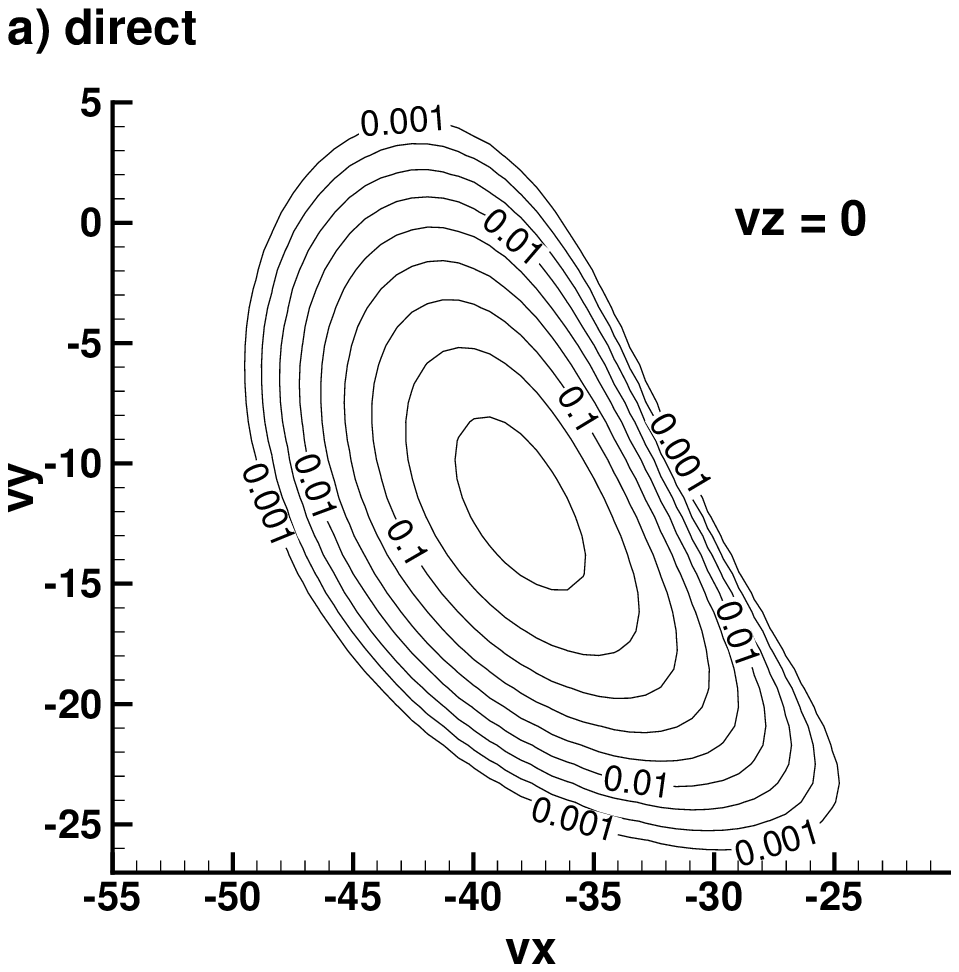}
\includegraphics[width=0.3\textwidth]{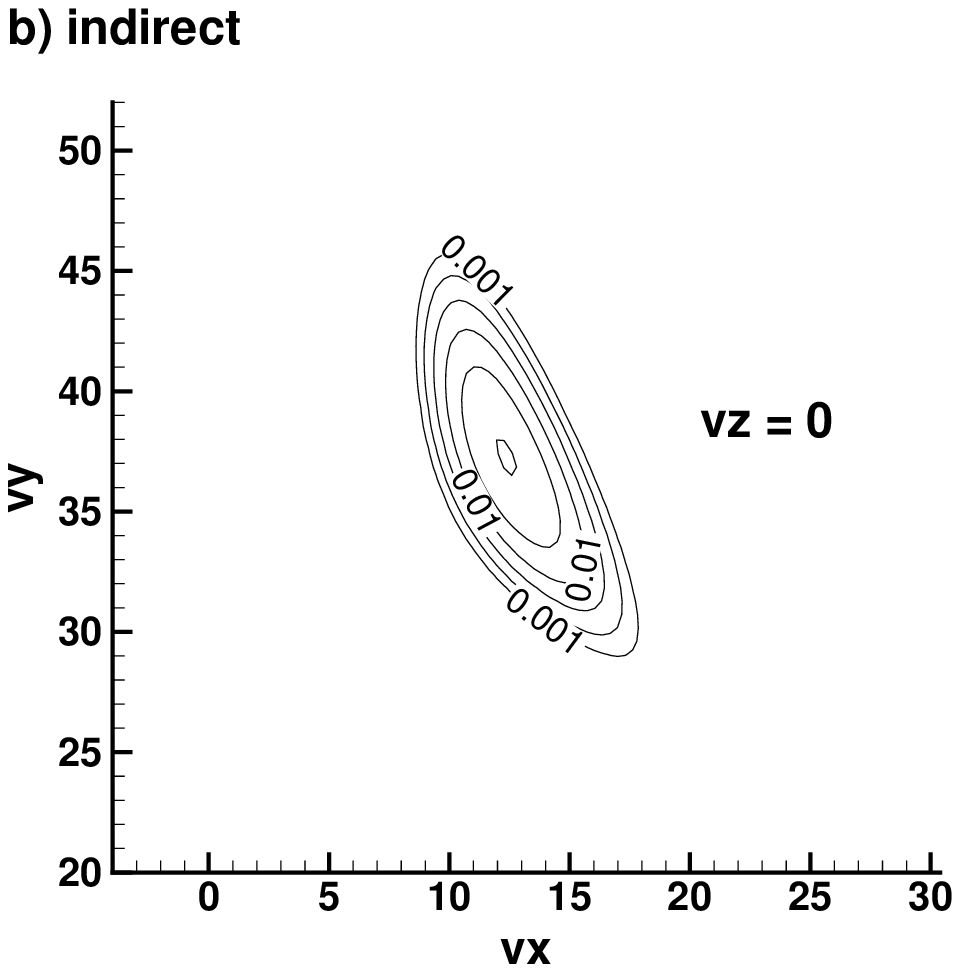}
\includegraphics[width=0.3\textwidth]{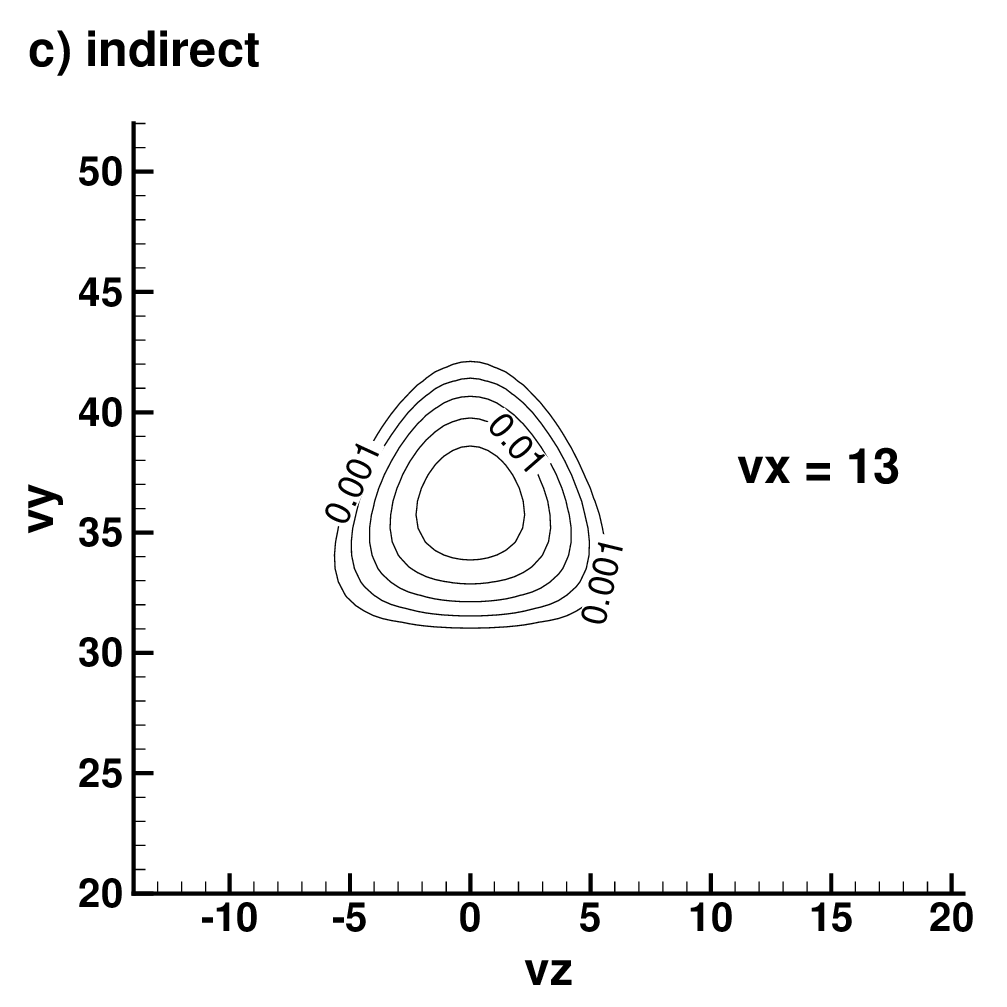}
\caption{Phase space density cuts of primary neutral interstellar
helium at a sidewind location (0, 2) through the direct (a) and
the indirect (b, c) PDF.  }
\end{figure}

Moving $\vec r_0$ slightly off the symmetry axis (Fig.\ 1c, d), it
can be seen that the degeneracy is broken. The plane of the torus
is now tilted with respect to the $v_y$-$v_z$ plane, and the torus
now connects a thick knot (maximum 62\%) to a small knot (34\%),
with the connections having only very low phase space density.
Going further off-axis makes it clear that the thick knot is
associated with direct path trajectories; the small knot is
associated with the indirect path (more prone to losses and hence
less dense), and they are no longer connected. Inspecting upwind
directions, the direct portion approaches a Maxwellian, whereas
the indirect portion is severely deformed, having to travel to the
Sun first and then back to the point of interest, with the
incumbent losses along the way.

An intermediate situation is shown in Figure 2, for the sidewind
location (0, 2). Direct and indirect PDF are well separated. The
direct peak (64\%) occurs for negative $v_x$ and negative $v_y$,
whereas both are positive for the indirect peak (5\%). The 3D
shapes are lenticular for the direct PDF, and even more irregular
for the indirect one. The velocity volume occupied by these PDF
traces back to the ISM and its Maxwellian. In configuration space,
all the associated trajectories go through the location (0, 2),
but at a remote reference distance, for example a sphere of radius
1000 AU, they fill a certain area. Note that not all interstellar
particles passing through that area intersect the location (0, 2);
in that sense, one can only talk about mapping the ISM Maxwellian
to the PDF at (0, 2) in velocity space, but not in configuration
space.

With the fast, accurate characterization of the neutral PDF in
hand, it is straightforward to perform integrals in velocity space
including moments of the PDF, to arrive at configuration space
maps of number density, bulk velocity, temperature, loss and
production rate, and others. All of these quantities also can be
obtained separately for direct and indirect neutrals. Given the
secondary production rate, through forward calculation the peak of
the secondary PDF at an arbitrary location $\vec r_0$ can be
constrained, making the calculation of the secondary PDF (scanning
a velocity neighborhood of the expected center) more efficient. As
mentioned above, for each individual secondary PDF point, an
integration (nominally from bow shock to heliopause) is necessary
to sum up the production terms and convolve the integration as
usual with losses on the way to $\vec r_0$. This integration
involves the knowledge of the primary PDF, and as such, the
process has a recursive flavor to it. As seen from the vantage of
heliospheric physics, one of the less resolved ingredients in the
problem of secondary PDF is how the helium ions necessary for
c.x.\ are being handled correctly. This problem is outside the
purview of the neutral solver exhibited in this contribution, but
pertains more to the correct handling of heavy ions through the
global plasma heliosphere models.

\section{Conclusions}

Using the Kepler equations of celestial mechanics is an elegant,
fast way to calculate trajectories of neutral heavy atoms
originating in the local interstellar medium. The associated
conserved quantities enable one-step calculations of particle
locations and velocities that are accurate by design. The
presented examples for helium show the sometimes complicated
heliospheric phase space distribution functions. They emphasize
the role of gravitational deflection and the effect of loss and
production processes on the helium phase space distribution. The
primary neutral helium PDF can be calculated in this way at any
arbitrary point in the heliosphere, and its various moments yield
maps of effective temperature, bulk velocity, and density. The
latter are elevated in the region of the helium focusing cone,
which is also where direct and indirect solutions merge together
and therefore become equally important.

The developed computational method, if applied only to primaries,
is somewhat parallel to the hot model of neutrals \citep{Fahr71}.
Yet the same method applies to secondary helium neutrals as well,
which is a novel feature. The method furthermore treats the plasma
as background and therefore can be used with a variety of global
heliospheric models to study the sensitivity of the helium signal
to heliospheric asymmetries or other differences in the global
models. Lastly, the mathematics of the method are not affected by
a time-dependence neither of the background plasma nor of the
photoionization rate; the method is therefore adaptable to
realistic, time-dependent scenarios, necessitating only an
increased level of housekeeping to match times in the trajectory
with the time-dependent inputs.

\acknowledgements Partial support of this work by NASA grants
NNX10AC44G, NNX11AB48G, NNX10AE46G, and by a University of Chicago
subcontract of NASA grant NNG05EC85C is gratefully acknowledged.
The author thanks Vladimir Florinski, Priscilla Frisch, and Gary
Zank for helpful interactions.


\hyphenation{Post-Script Sprin-ger}

\end{document}